\pdfoutput=1

\documentclass[aip,jcp,reprint,floatfix,citeautoscript]{revtex4-1}

\pdfpageattr {/Group << /S /Transparency /I true /CS /DeviceRGB>>}
\usepackage{cmap}
\usepackage[T1]{fontenc}
\usepackage{helvet}

\usepackage[UKenglish]{babel}
\usepackage{csquotes}

\usepackage{newtxtext,newtxmath}
\DeclareSymbolFont{UPM}{U}{eur}{m}{n}
\DeclareMathSymbol{\uppartial}{0}{UPM}{"40}

\usepackage{booktabs}
\usepackage{multirow}

\usepackage{xparse}
\def\rowwidth{*}
\newcommand\splitEntry[1]{\headersep{}\multirow{-2}{\rowwidth}{#1{}}\def\headersep{&}}
\DeclareDocumentCommand\rowentry{>{ \SplitList{;}} m }{\def\rowwidth{3.5cm}\def\headersep{\!}\relax\ProcessList{#1}{\splitEntry}\relax}

\usepackage[dvipsnames, table]{xcolor}

\usepackage{xifthen}

\usepackage[pdftex, bookmarks, pdffitwindow=false, pdfstartview=FitH, pdfdisplaydoctitle, colorlinks, plainpages=false, pdftitle={Phase behaviour of empirical potentials of titanium dioxide},pdfauthor={Aleks Reinhardt}, pdfpagelabels, hypertexnames, citecolor={cyan},linkcolor={cyan}, urlcolor={violet}, pdflang={en}, hyperfootnotes=false, breaklinks]{hyperref}
\usepackage[all]{hypcap}

\usepackage{microtype}

\usepackage{graphicx}
\usepackage{amsmath}

\usepackage{flafter}
\usepackage[version=3,arrows=pgf]{mhchem}
\usepackage{siunitx}   %% to make typesetting numbers and units easier
\usepackage[italic,basic]{mathastext}

\usepackage{mleftright}

\newcommand{\der}{\ensuremath{\mathrm{d}}}

\usepackage[nodayofweek]{datetime}
\newdateformat{myDate}{\THEDAY\ \monthname[\THEMONTH] \THEYEAR}

\newcommand{\refSub}[2]{\hyperref[#2]{\ref{#2}(#1)}}

\begin{document}

\title{Phase behaviour of empirical potentials of titanium dioxide}\thanks{Published in J.~Chem.~Phys.~\textbf{151}, 064505 (2019), \href{https://doi.org/10.1063/1.5115161}{doi:10.1063/1.5115161}.}
\author{Aleks Reinhardt}
\affiliation{Department of Chemistry, University of Cambridge, Lensfield Road, Cambridge, CB2 1EW, United Kingdom}
\date{\myDate\today}

\raggedbottom

\begin{abstract}
In recent years, several relatively similar empirical models of titanium dioxide have been proposed as reparameterisations of the potential of Matsui and Akaogi, with the Buckingham interaction replaced by a Lennard-Jones interaction. However, because of the steepness of the repulsive region of the Lennard-Jones potential, such reparameterised models result in rather different mechanical and thermodynamic properties compared to the original potential. Here, we use free-energy calculations based on the Einstein crystal method to compute the phase diagram of both the Matsui--Akaogi potential and one of its Lennard-Jones-based reparameterisations. Both potentials are able to support a large number of distinct crystalline polymorphs of titanium dioxide that have been observed in experiment, but the regions of thermodynamic stability of the individual phases are significantly different from one another. Moreover, neither potential results in phase behaviour that is fully consistent with the available experimental evidence.
\end{abstract}

\maketitle

\section{Introduction}
Titanium dioxide is a material with many applications, and it is therefore perhaps not overly surprising that a number of empirical potentials\cite{Matsui1991, Kim1996, Pedone2006, Luan2015, Alderman2014, Brandt2015} that enable computer simulations to be tractable have been developed. Although some of the material's most interesting behaviours, such as its photocatalytic activity and chemical reactivity, are surface-driven effects, for which quantum and electromechanical effects are crucial to consider properly~\cite{Cheng2014, Predota2004, Liu2010, LamielGarcia2017}, titania is used in paints and sunscreens due to its interesting optical properties, which largely depend on the size of the titania particles and their interaction with the solvent. Simpler potentials may be sufficient to describe the behaviour of TiO$_2$ nanoparticles in aqueous solution or their interactions with some organic molecules~\cite{YazdanYar2018}, the bulk phase behaviour of TiO$_2$ solids~\cite{Shang2014}, and can be useful in preparing initial nanoparticle structures for subsequent analysis with more complex models~\cite{MaciaEscatllar2019}. In fact, in some cases, simple models of titanium dioxide outperform more complex ones~\cite{Swamy2001, Naicker2005}.

One of the most widely used empirical potentials for the condensed phases of TiO$_2$ is the Matsui--Akaogi (MA) potential~\cite{Matsui1991}, which reproduces many experimental features of titania well~\cite{Collins1996, Futera2016}. It comprises a Coulomb interaction term and a Buckingham potential term to account for the atoms' excluded volume. The MA potential has been extended over the years in various ways, for example to describe TiO$_2$--water\cite{Bandura2003} and TiO$_2$--ion  interactions~\cite{Predota2004} or lithiated titania~\cite{Kerisit2010}. Several authors have also reparameterised the Buckingham part of the MA potential in terms of the Lennard-Jones (LJ) potential~\cite{Kang2010, Friedrichs2014, Luan2015, Brandt2015} with the aim of facilitating simulations of TiO$_2$ surfaces and surface thermodynamics~\cite{Friedrichs2014, Brandt2015}, TiO$_2$ in aqueous solvents~\cite{Friedrichs2014}, TiO$_2$ nanoparticles\cite{Luan2015, Brandt2015} or interactions of TiO$_2$ with biological molecules such as proteins or biomembranes~\cite{Kang2010, Friedrichs2014, Brandt2015, Luan2015}. However, each of these reparameterisations is slightly different, and so a large number of very similar -- but not quite identical -- empirical models of TiO$_2$ are now available.

In this work, we compare the thermodynamic behaviour of the MA potential and one of its several reparameterisations using the Lennard-Jones potential. By explicitly calculating the free energy of a number of the crystalline polymorphs of TiO$_2$, we demonstrate which phases are thermodynamically stable under changing conditions of pressure and temperature. In addition to the most familiar phases of TiO$_2$, rutile and anatase, empirical potentials are capable of describing a wide range of other observed and hypothesised polymorphs. The calculation of phase diagrams is a relatively laborious task, but it can provide useful insight into the behaviour of a particular model. By calculating the free energies of an array of crystal structures of not only the MA potential of TiO$_2$, but also of one of the reparameterised forms of the potential, we demonstrate that seemingly small changes in the potential can have very significant implications for thermodynamic stability and metastability and for the phase behaviour of a system.

\section{Reparameterisations of the Matsui--Akaogi potential}
\begin{table}[t!]
\caption{Parameters for the Matsui--Akaogi potential\cite{Matsui1991} ($A_{ij}$, $B_{ij}$ and $C_{ij}$) and the Luan--Huynh--Zhou potential\cite{Luan2015} ($\varepsilon_{ij}$ and $\sigma_{ij}$).}\label{tab:MApot}
\centering
% \figureversion{lf,tab}
\sisetup{detect-weight=true, detect-family=true, detect-mode=true}%,text-rm={\figureversion{tab,lf}}}
\begin{tabular}{c c @{\hspace{0.7em}}S[table-format=5.1] @{\hspace{0.7em}}S[table-format=1.3] @{\hspace{0.7em}}S[table-format=2.3]  @{\hspace{.7em}}S[table-format=1.3] @{\hspace{0.7em}}S[table-format=1.3]}\toprule
$i$ & $j$ & $A_{ij}/\si{\electronvolt}$ & $B_{ij}/\si{\angstrom}$ & {$C_{ij}/(\si{\electronvolt\angstrom\tothe{6}})$} & {$\varepsilon_{ij}/(\SI{e-2}{\electronvolt})$} & $\sigma_{ij}/\si{\angstrom}$ \\\midrule
Ti & Ti & 31120.4 & 0.154 & 5.247 & 2.515 & 1.960\\
Ti & O & 16957.7 & 0.194 & 12.593 & 1.839 & 2.423 \\
O & O & 11782.8 & 0.234 & 30.222 & 1.344 & 2.887\\
\bottomrule
\end{tabular}
\end{table}

The functional form of the MA potential~\cite{Matsui1991} is a sum of a Coulomb term (with charges $q_{\text{Ti}}=2.196e$ and $q_{\text{O}}=-1.098e$, where $e$ is the elementary charge) and the Buckingham potential~\cite{Buckingham1938}
\begin{equation}
 \phi_\text{Buck}(i,\,j,\,r_{ij}) = A_{ij} \exp(-r_{ij}/B_{ij}) - C_{ij} r_{ij}^{-6},\label{eqn-buck}
\end{equation}
where $r_{ij}$ is the distance between atoms $i$ and $j$, and the parameters $A_{ij}$, $B_{ij}$ and $C_{ij}$ are given in Table~\ref{tab:MApot}.

Like any locally quadratic functions, the Lennard-Jones and Buckingham potentials look similar around their minima.
The form of the Buckingham potential is somewhat more physically motivated than that of the Lennard-Jones potential~\cite{VanVleet2016}.
However, the Buckingham potential is not implemented in some popular simulation software packages (such as \textsc{Amber})~\cite{Luan2015}, and so several groups have tried to parameterise a potential analogous to the MA potential, but with the Buckingham potential replaced by the Lennard-Jones potential.
Whilst generalised Lennard-Jones potentials can offer a good approximation to the Buckingham potential~\cite{Lim2009}, the way this reparameterisation was largely achieved in previous work was to consider the Buckingham and Lennard-Jones potentials without accounting for the Coulomb interaction, and then finding the best set of fitting parameters to map the former onto the latter~\cite{Kang2010, Friedrichs2014, Luan2015, Brandt2015}.

However, as two nuclei approach one another, the repulsive $r^{-12}$ Lennard-Jones term quickly becomes very much steeper than the exponential term in the Buckingham potential~\cite{Kang2010}.
If the Lennard-Jones fitting parameters are computed around the minimum of the potential, the resulting potential will thus of necessity be considerably poorer at smaller internuclear separations.
Unfortunately, this is precisely the region that is most important to capture well within the Matsui--Akaogi model of TiO$_2$, since if we add the Coulomb potential to either the Lennard-Jones or the Buckingham potentials, for oppositely charged particles, the local minimum in the overall combined potential shifts to much smaller interparticle distances, into the region where the Lennard-Jones potential is much more steeply -- and unphysically~\cite{VanVleet2016} -- repulsive than the Buckingham potential.
Indeed, the minimum due to dispersion ($r^{-6}$) interactions is essentially negligible compared to the Coulomb interaction in the MA potential, and the crucial behaviour to describe in this parameterisation is therefore the Pauli repulsion.
As a result, the existing LJ-based reparameterisations of the MA potential do not reproduce the important parts of the potential energy landscape particularly well, and are likely to result in behaviour that is rather different from the MA potential.

Although the MA potential has been described as being equivalent to certain of its Lennard-Jones reparameterisations in some previous work~\cite{Brandt2015}, the densities of the crystalline phases are in fact found to be significantly different~\cite{Brandt2015, Luan2015}.
This is not surprising in the light of the LJ potential's steeper repulsion in the region where the overall MA potential is minimised.
The densities of the fluid phase are also significantly different (of the order of \SI{10}{\percent}), which will almost certainly significantly affect the nature of the structure of the amorphous nanoparticles investigated by Luan~\textit{et al.}~\cite{Luan2015}
Seemingly trivial reparameterisations of empirical potentials can thus result in very significant differences in the outcome, but no systematic investigation into the phase behaviour of these various empirical models of titanium dioxide seems to have been undertaken thus far.

While the various Lennard-Jones reparameterisations of the MA potential reported thus far are not completely identical, the differences between them are small. Here, we therefore use the Luan--Huynh--Zhou (LHZ) potential of Ref.~\onlinecite{Luan2015} as a representative of these reparameterised potentials. In the LHZ potential, the Coulomb term is identical to the MA potential, but the Buckingham potential is replaced by the 12--6 Lennard-Jones potential~\cite{Jones1924}
\begin{equation}
  \phi_\text{LJ}(i,\,j,\,r_{ij}) = 4\varepsilon_{ij} \left[\left(\sigma_{ij}/r_{ij}\right)^{12} - \left(\sigma_{ij}/r_{ij}\right)^{6}\right],\label{eqn-lj}
\end{equation}
where $\varepsilon_{ij}$ is the well depth and $\sigma_{ij}$ is the effective particle diameter, with parameters given in Table~\ref{tab:MApot}~\cite{Note1}.

\section{Phase diagrams and free-energy calculations}

Phase transformations of TiO$_2$ have been studied in nano\-rods and nano\-tubes with the MA model~\cite{Buin2011}, as has anatase--rutile nucleation in nanoparticles~\cite{Zhou2012}.
The phase diagram at absolute zero in temperature can readily be computed in simulations by finding the minimum of the potential energy.
Such phase diagrams have been reported for empirical potentials using energy minimisation~\cite{Swamy2001}, and more generally, a number of density-functional theory (DFT) and quantum Monte Carlo calculations have been used to approximate the phase diagram at absolute zero~\cite{Muscat2002, Ma2009, Dekura2011, Swamy2014, Liu2015, Luo2016, Trail2017}.
For phases with relatively similar energies, entropy differences between the phases can determine the phase behaviour~\cite{Fu2013}.
Some attempts have been made to determine the phase diagram at finite temperatures from DFT calculations\cite{Yu2009, Hu2010, Fu2013, Mei2014, Luo2016} by using e.g.~the (quasi-harmonic) Debye model\cite{Blanco2004} with phonon dispersion relations to approximate the free energy of the system as a function of temperature.

Here, we use direct free-energy calculations\cite{Frenkel1984, Vega2008, Aragones2012, Aragones2013} to determine the phase diagram of the crystalline forms of TiO$_2$ for the MA and LHZ potentials. In particular, we use the Frenkel--Ladd approach\cite{Frenkel1984, Vega2008} to calculate the free energy of a crystal of interest by thermodynamic integration from an Einstein crystal, a reference state for which the free energy is known from statistical mechanics.
The procedure is as follows: we equilibrate a crystalline phase of interest in an $NPT$ simulation to determine the equilibrium lattice parameters at a temperature and pressure at which the phase is stable or metastable. We then take an ideal crystal, constructed from those lattice parameters, and tether each atom to its lattice site by a harmonic spring. We calculate the Helmholtz energy of the corresponding Einstein crystal analytically, and then use hamiltonian thermodynamic integration to switch on the potential of interest. Finally, we use hamiltonian thermodynamic integration to compute the Helmholtz energy $A$ of the state of interest as we switch off the harmonic springs. The details of the procedure, including a discussion of the need to keep the overall centre of mass of the system fixed, are described in Ref.~\onlinecite{Vega2008}.

At phase equilibrium, two phases must have equal temperatures, pressures and chemical potentials, and so it is helpful to compute their chemical potentials as the Gibbs energies per particle, $\beta\mu = \beta G/N = \beta A/N + \beta P/\rho$, where $\beta=1/k_\text{B}T$, $N$ is the number of atoms in the system, $A$ is the Helmholtz energy, $P$ is the pressure, and $\rho=N/V$ is the number density.
Once the chemical potential is known at a given point, we can use standard thermodynamic integration in pressure or temperature to determine the chemical potential under different thermodynamic conditions.
For example, we can integrate the Gibbs--Duhem relation, $\der \mu=v\,\der P - s\,\der T$, at constant temperature to give
\begin{equation} \mu(P) = \mu(P_0) + \int_{P_0}^P v(P')\,\der P',\end{equation}
or, we can integrate the Gibbs--Helmholtz equation, $(\uppartial  (G/T)/\uppartial T)_{N,\,P}=-H/T^2$, at constant pressure to give
\begin{equation}
\beta \mu(T) = \beta_0 \mu(T_0) - \int_{T_0}^{T} \frac{H}{Nk_\text{B}(T')^2}\,\der T',
\end{equation}
where $H$ is the enthalpy~\cite{Note2}.

In order to compute coexistence points, we first determine chemical potentials of different phases at a given temperature and a specified pressure.
We then perform thermodynamic integration along an isotherm to obtain the chemical potential of each phase at this temperature as a function of pressure.
At the pressure at which the chemical potentials of different phases intersect, these phases are at equilibrium with one another, and this point corresponds to phase coexistence.

Once a coexistence point is known for a pair of phases, we can trace it across the phase diagram by numerically integrating the Clapeyron equation,
\begin{equation}  \frac{\der P}{\der T} =  \frac{\upDelta_\text{trs} H}{T\upDelta_\text{trs} V},  \label{eqn-clapeyron}\end{equation}
a procedure known as Gibbs--Duhem integration~\cite{Vega2008, Kofke1993, *Kofke1993b}.
In solid--solid phase transitions, the effect of changing the pressure is likely to be more significant than that of changing the temperature, and so $\der P/\der T$ will be relatively small, resulting in a reasonably stable numerical integration.
We use a fourth-order Runge--Kutta algorithm to perform it.
However, whilst Gibbs--Duhem integration is a convenient method, it lacks any error-checking mechanism: small numerical errors can quickly balloon and the predicted coexistence line can deviate from the true coexistence line.
One should therefore not rely on the method in isolation, particularly over large temperature ranges.
More advanced approaches, such as free-energy extrapolation~\cite{Escobedo2014}, can be used, but here we use Gibbs--Duhem integration only as a check of independent free-energy calculations at other temperatures and pressures.

To obtain chemical potentials at other temperatures, we can either perform a new set of Frenkel--Ladd calculations, or start from the chemical potential we obtained at one temperature and integrate it along an isobar to the temperature of interest. Here, we use both approaches: provided they give results in agreement with one another, we can have a degree of confidence that the data are robust. In both cases, we obtain the chemical potential at a given temperature and pressure, and we can then proceed as above by integrating along an isotherm to determine the chemical potential of each phase of interest as a function of pressure in order to determine points of coexistence.

Although in principle, we could also investigate the melting behaviour of the various crystal polymorphs studied here by computing the Gibbs energy of the fluid phase by thermodynamic integration from a suitable perfect gas, the TiO$_2$ melt has been shown not to be very well represented by the MA potential~\cite{Alderman2014}, and so we do not focus on the very high temperature behaviour of the system here.

\begin{table*}
\caption{Representative densities and unit cell lattice parameters of crystal structures considered are given for both potentials used. Samples of each phase considered were equilibrated at \SI{700}{\kelvin} and the pressure specified below.
Initial configurations were constructed from the crystallographic data provided in the reference given in each case.}\label{table:phaseEql}
\centering
% \figureversion{lf,tab}
\sisetup{detect-weight=true, detect-family=true, detect-mode=true}%,text-rm={\figureversion{tab,lf}}}
\newcolumntype{Z}{S[table-format=2.3(2)]}
\def\colRw{\rowcolor{black!10}[\tabcolsep]}
\sisetup{table-format=2.2}
\small
\begin{tabular}{l >{\footnotesize}c c c *5{Z}}
\toprule
Phase & { \parbox{1cm}{Space group}} & {$P/\si{\giga\pascal}$} & Model & $\rho/\si{\gram\per\cubic\centi\metre}$ & $a/\si{\angstrom}$ & $b/\si{\angstrom}$& $c/\si{\angstrom}$& $\beta/\si{\degree}$ \\\midrule
\colRw & & &   MA  & 3.851(5) & 3.777(3) &   & 9.653(11) & \\
\colRw\rowentry{anatase~\cite{Rezaee2011} ; I4$_1$/\textit{amd} ; \tablenum{0.75}} & LHZ & 3.396(4) & 3.950(1) & & 10.010(7) & \\
& & &   MA & 4.131(6) & 9.188(13) & 5.406(6) & 5.171(6) & \\
\rowentry{brookite~\cite{Meagher1979} ; P\textit{bca} ; \tablenum{0.75} } & LHZ & 3.657(5) & 9.638(10) & 5.624(7) & 5.353(7) & \\
\colRw& & &   MA & 3.667(4) & 12.347(12) & 3.752(4) & 6.517(7) & 106.6(1)  \\
\colRw\rowentry{TiO$_2$(B)~\cite{Marchand1980} ; C2/\textit{m} ; \tablenum{0.75} } & LHZ & 3.226(4) & 12.881(12) & 3.928(3) & 6.723(7) & 104.8(1) \\
& & &   MA & 4.312(5) & 4.507(3) &  & 3.028(3) & \\
\rowentry{rutile~\cite{Rezaee2011} ; P4$_2$/\textit{mnm} ;  \tablenum{0.75} } & LHZ &  3.826(4) & 4.686(2) &  & 3.158(3)\\
\colRw& & &   MA & 3.487(5)  & 9.998(7) &  & 3.044(3) & \\
\colRw\rowentry{hollandite (TiO$_2$(H))~\cite{Sasaki1993} ; I4/\textit{m} ; \tablenum{4} } & LHZ &  3.079(5) & 10.413(7) &  & 3.178(3) & \\
& & &   MA & 4.536(4) & 4.489(5) & 5.325(5) & 4.892(4) & \\
\rowentry{columbite (TiO$_2$-II)~\cite{Grey1988} ; P\textit{bcn} ; \tablenum{10} } & LHZ & 4.026(3) & 4.726(6) & 5.484(6) & 5.084(5) & \\
\colRw& & &   MA & 4.938(5) & 4.679(4) & 4.831(3) & 4.780(4) & 96.15(8) \\
\colRw\rowentry{baddeleyite~\cite{Swamy2005} ; P2$_1$/\textit{c} ; \tablenum{20} } & LHZ & 4.477(4) & 4.846(4) & 5.036(3) & 4.889(3) & 96.75(8)\\
 & & &   MA & 5.052(7) & 4.718(2)& & & \\
\rowentry{pyrite~\cite{Zhu2014}  ; P\textit{a}$\overline{\text{3}}$ ; \tablenum{25} } & LHZ & 4.493(6) & 4.906(2) & & & \\
\colRw& & &   MA & 5.167(4) & 9.193(9) & 4.791(4) & 4.662(4) &\\
\colRw\rowentry{OI~\cite{Dubrovinskaia2001} analogue ; P\textit{bca} ; \tablenum{30} } & LHZ & 4.640(4) & 9.507(9) & 4.980(4) & 4.830(4) & \\
& & &   MA & {not stable} & & & & \\
\rowentry{oxygen-displaced fluorite~\cite{Zhou2010}  ; P\textit{ca}2$_1$ ; \tablenum{35} }& LHZ & 4.667(4) & 4.934(5) & 4.789(4) & 4.809(4) & \\
\colRw& & &   MA & 5.786(5) & 5.093(5) & 3.055(4) & 5.893(6) & \\
\colRw\rowentry{cotunnite (OII)~\cite{Dubrovinsky2001} ; P\textit{nma} ; \tablenum{55} } & LHZ & 5.211(2) & 5.241(3) & 3.158(2) & 6.151(3) & \\
& & &   MA & 5.481(7) & 4.591(2) & & & \\
\rowentry{fluorite  ; F\textit{m}$\overline{\text{3}}$\textit{m} ; \tablenum{60}}  & LHZ & 4.887(5) & 4.770(2) & & & \\
\bottomrule
\end{tabular}
\end{table*}

\section{Simulation details}
Molecular dynamics simulations using the potentials introduced above were performed with the \textsc{Lammps} simulation package~\cite{Plimpton1995} (v.~16Mar18) within an overarching free-energy calculation code. The Buckingham part of the MA potential was truncated at a cutoff of \SI{12}{\angstrom}. As in Ref.~\onlinecite{Luan2015}, a smoothing function was used for the Lennard-Jones part of the LHZ potential between \SI{10}{\angstrom} and \SI{12}{\angstrom}. To account for long-range electrostatics in both the MA and the LHZ cases, PPPM summation~\cite{Hockney1988} with a real-space cutoff of \SI{12}{\angstrom} was used. In brute-force equilibration simulations, the integration time step in the velocity Verlet algorithm was \SI{1}{\femto\second}, while in Frenkel--Ladd simulations, a shorter time step of \SI{0.1}{\femto\second} was necessary to maintain stability. In most simulations where the temperature or pressure needed to be conserved, a Nos\'e--Hoover thermostat~\cite{Nose1984, Hoover1985} and a Parrinello--Rahman barostat were used~\cite{Parrinello1981}. However, since the  Nos\'e--Hoover thermostat does not reproduce the canonical distribution for harmonic motion~\cite{Hoover1985, Aragones2013}, in Einstein crystal simulations involving harmonic springs, velocity rescaling with the Bussi thermostat\cite{Bussi2007} was used instead.

Finite size effects can be appreciable in free-energy calculations of solids~\cite{Vega2008}. We use systems of the order of 3500 particles, where the finite-size correction for the chemical potential is estimated to be less than $0.005\,k_\text{B}T$, and so we can ignore it. Sufficiently large system sizes are essential in this case, since the gradients of the free energy of some of the phases are so similar that a small error in the calculation of the free energy at a given starting point can lead to a large error in the coexistence point following thermodynamic integration.

Determining absolute chemical potentials from simulations entails a number of simulation methods used in combination, each of which produces results with a certain error bar. Propagating errors is not straightforward, and so, as suggested by Vega~\textit{et al.}~\cite{Vega2008}, we perform consistency checks at the end of the calculation. For all phases with free energies that could be competitive, we compute the Gibbs energies at a given temperature using Frenkel--Ladd integration at a range of different pressures, and make sure that data obtained in this way match up when using thermodynamic integration along an isotherm to the same point from one of the other data points obtained from the Frenkel--Ladd approach. At any given temperature, this leads to errors in free energies that are smaller than $0.01\,k_\text{B}T$ per particle. Similar errors are obtained when performing thermodynamic integration in temperature along isobars~\cite{Note3}, even when integrating over several hundred kelvin, with chemical-potential differences between end-points of the order of $80\,k_\text{B}T$.  However, it is worth bearing in mind that such errors, while seemingly small, can nevertheless shift coexistence points by an appreciable fraction.

\section{Results and discussion}

\begin{figure*}[t]
\centering
\includegraphics{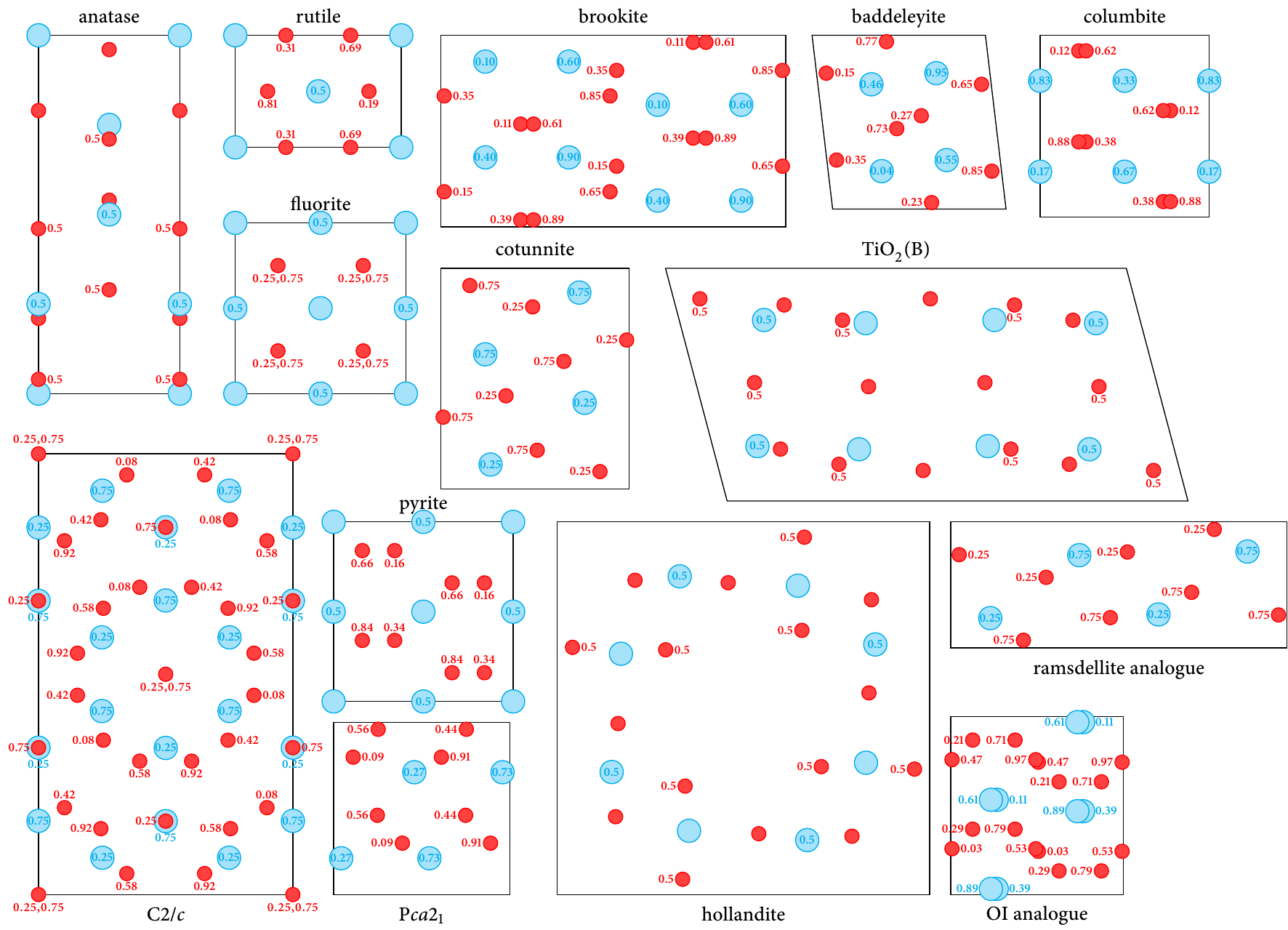}
\caption{Plan views of unit cells of the crystal structures considered, as labelled. In all cases except for hollandite and C/2\textit{c}, the view is down the $b$ axis (with dimensions corresponding to Table~\ref{table:phaseEql}). In the case of hollandite and C2/\textit{c}, for clarity, the view is along the $c$ axis. Relative heights above the plane are indicated as fractions of the lattice parameter along the projected axis; where no height is indicated, this corresponds to a relative height of 0 and 1. Titanium atoms are shown in cyan and oxygen atoms in red. While relative distances are consistent across all structures shown, they correspond to different pressures, and thus serve only to give a rough indication of the packing density. Full crystal structures are available in the supporting data.}
\label{fig-mainPhases}
\end{figure*}

\begin{figure*}[t]
\centering
\includegraphics{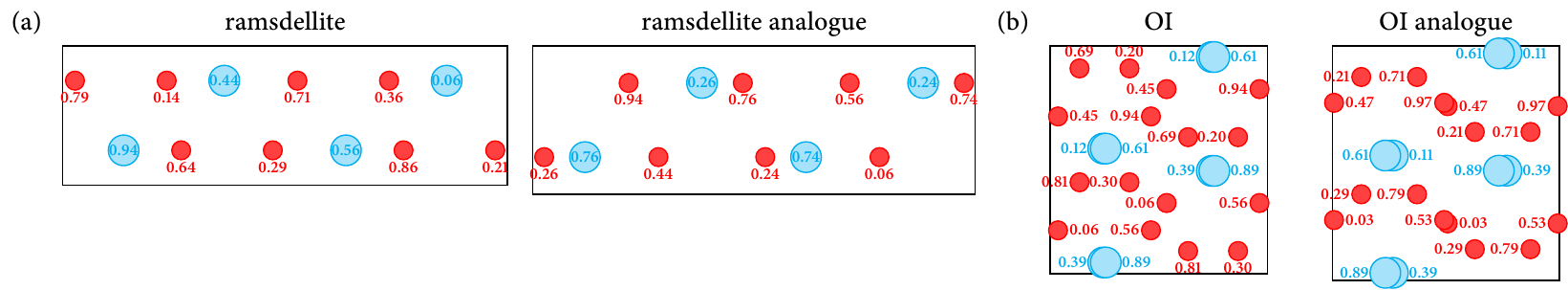}
\caption{(a) Plan views of ramsdellite from Ref.~\onlinecite{Akimoto1994} and the structure from simulation with the MA potential; both belong to the P\textit{bnm} (P\textit{nma}) space group. The structure is not stable with the LHZ potential and converts to the OI analogue. (b) Plan views of the OI phase from Ref.~\onlinecite{Dubrovinskaia2001} and the structure from simulation (the structure is essentially the same with either the MA or the LHZ potential). Both belong to the P\textit{bca} space group. Full crystal structures are available in the supporting data.}
\label{fig-additionalPhases}
\end{figure*}

Starting structures were initially equilibrated in $NPT$ simulations using both potentials under conditions roughly corresponding to the conditions at which the phases were reported in the literature. For an illustration of the widely different densities and unit cell parameters among the different phases, representative equilibrated data are shown in Table~\ref{table:phaseEql}.

In addition to the phases listed in this table, and illustrated in Fig.~\ref{fig-mainPhases}, which have previously been reported or hypothesised for TiO$_2$, several additional phases were obtained in simulations. For example, cotunnite, when simulated with the MA potential, converts to the very open C2/\textit{c} structure [illustrated in Fig.~\ref{fig-mainPhases}] below approximately \SI{10}{\giga\pascal}. The phase is not thermodynamically stable, but it is kinetically accessible, as it was obtained from brute-force simulations. Moreover, two phases reported in the literature for TiO$_2$, ramsdellite~\cite{Akimoto1994} and TiO$_2$(OI)~\cite{Dubrovinskaia2001}, have different motifs when simulated using the empirical potentials considered here [see Fig.~\ref{fig-additionalPhases}].

\begin{figure*}[t]
\centering
\includegraphics{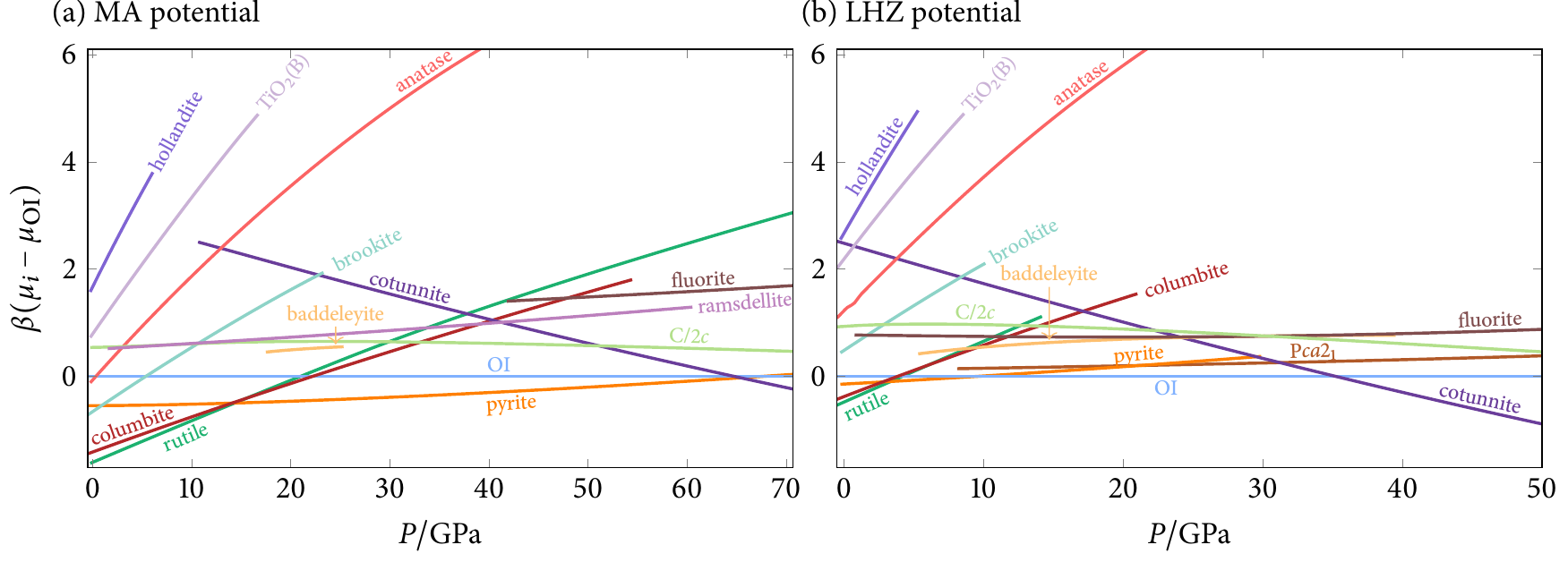}
\caption{Chemical potentials for the (a) MA and (b) LHZ potential at \SI{700}{\kelvin} as a function of the pressure, shown relative to that of the OI phase. Note that the pressure axis spans a smaller pressure range in panel (b).}\label{fig-chemPot-both}
\end{figure*}

\begin{figure*}[t]
\centering
\includegraphics{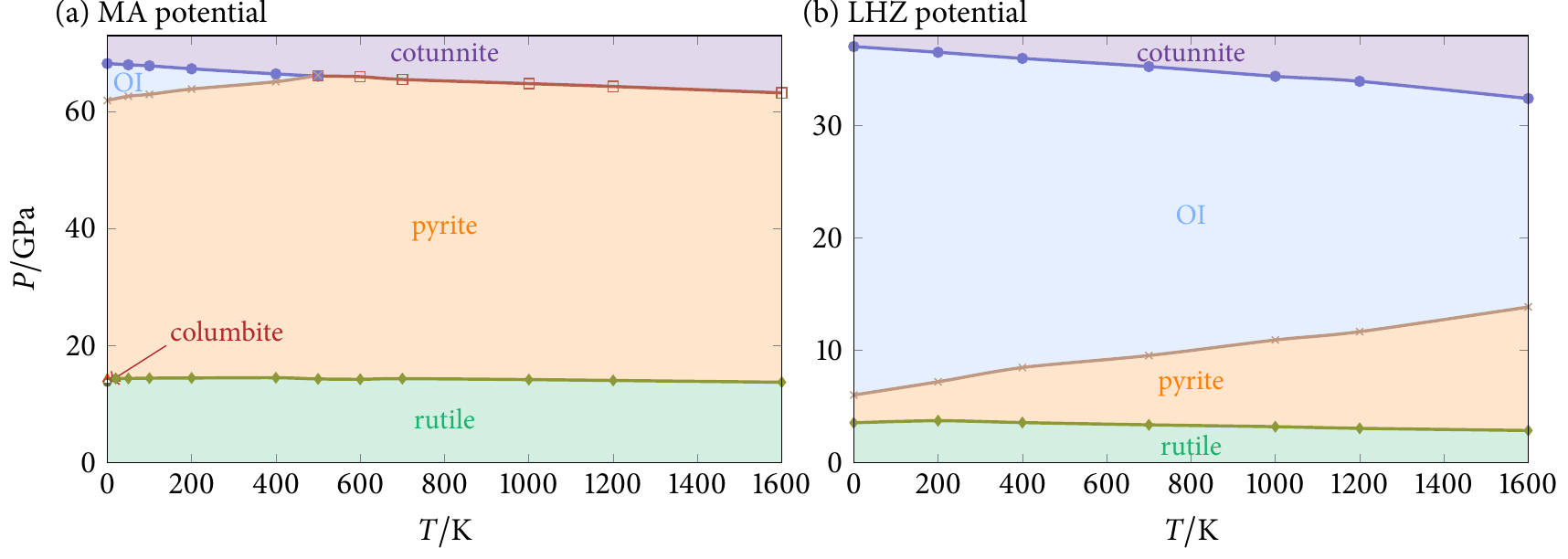}
\caption{Pressure--temperature phase diagrams for (a) the MA and (b) the LHZ potential. Each marker corresponds to a full free-energy calculation; the markers are connected by straight lines, which are merely guides to the eye. Note that the pressure axis spans a smaller pressure range in panel (b).}\label{fig-phaseDiag-both}
\end{figure*}

In order to determine the chemical potentials of the structures, the Einstein crystal method was used, starting from energy-minimised perfect crystals with unit cell parameters first obtained in $NPT$ simulations, and then applying the Frenkel--Ladd formalism to obtain the free energy of the potential of interest.
Once the Gibbs energy of a phase was determined at one pressure and temperature, it was computed at other pressures by means of thermodynamic integration.
The resulting chemical potentials, as a function of pressure, are shown for a variety of phases in Fig.~\ref{fig-chemPot-both}.
For those phases that have a sufficiently low chemical potential that they could potentially be thermodynamically stable over some region of pressures, namely rutile, columbite, pyrite, the OI analogue, cotunnite and, in the case of the LHZ potential, the P\textit{ca}2$_1$ phase, the curves shown in Fig.~\ref{fig-chemPot-both} were obtained by computing chemical potentials in independent free-energy calculations at between 5 and 8 different pressures, using appropriate thermodynamic integration along the isotherm, and averaging the result as a function of pressure. The spread of the chemical potential obtained in this way was always less than $0.01\,k_\text{B}T$, which is better than the uncertainty of $0.05\,k_\text{B}T$ typically expected in such free-energy calculations~\cite{Vega2008}.

The curves shown in Fig.~\ref{fig-chemPot-both} correspond to the region over which the phases indicated are either thermodynamically stable or metastable in brute-force MD simulations.
Not all curves extend throughout the entire pressure range because the corresponding structures undergo a phase transition to a different polymorph, although, as we discuss below, the resulting polymorph is not always the thermodynamically stable phase.
The majority of the polymorphic phases of TiO$_2$ that have been reported in the literature can be simulated with either of the empirical potentials used, which suggests that these potentials are in some sense surprisingly versatile and sufficiently powerful to be able to represent a multitude of (metastable) polymorphs in computer simulation.
However, many of these polymorphs are relatively high in free energy, and so would not appear in an equilibrium phase diagram of the system.
Only the phases discussed above were considered in these free-energy calculations; there may of course be other phases which we have not considered because they have not (yet) been reported for TiO$_2$, and some of them may well have a lower free energy still for these empirical potentials.
It is not possible to say with certainty either in experiment or in simulations that a given phase really is thermodynamically stable; we can only make comparisons between phases of which we are aware and not absolute predictions.

Pressure--temperature phase diagrams, showing the phase with the lowest free energy under the conditions of interest, are shown in Fig.~\ref{fig-phaseDiag-both} for the two potentials.
These phase diagrams were determined directly from free-energy calculations by constructing analogues of Fig.~\ref{fig-chemPot-both} at a range of temperatures and determining the points at which chemical-potential curves of different phases cross over, and finding the phase with the minimum chemical potential as a function of pressure.
The results were also verified by performing Gibbs--Duhem integration as a function of temperature from several starting points and ensuring that independent integrations match up, which helps to confirm that the data shown are reasonably robust.

From Fig.~\ref{fig-phaseDiag-both} it is apparent that the phase diagrams corresponding to the LHZ and MA potentials are rather different.
The MA potential was parameterised to provide a reasonable match to the experimental density of rutile; by contrast, because the LHZ potential was reparameterised based solely on the form of the Buckingham part of the MA potential, without accounting for the fact that the true minimum in the potential energy was rather deep within the repulsive region, the densities of the MA polymorphs are approximately \SI{10}{\percent} to \SI{15}{\percent} larger than those of the LHZ analogues.
Since the MA potential is less steeply repulsive than the LHZ analogue at small interparticle distances, it is not perhaps too surprising that its solid phases are able to accommodate significantly higher pressures.
However, the various polymorphs are not uniformly affected by replacing the Buckingham potential with the LJ potential, and the thermodynamic phase behaviour of the two potentials is rather different.
The OI analogue, for example, is considerably more stable with the LHZ potential at higher temperatures, and so takes up the majority of the phase diagram shown in Fig.~\refSub{b}{fig-phaseDiag-both}, while the pyrite phase is stable over a broader range of pressures for the MA potential.

It is well known that seemingly small differences in internuclear potentials can result in markedly different phase behaviour; the most famous and well-studied example is perhaps water, where a number of similar empirical potentials exist, and yet their phase diagrams are remarkably dissimilar~\cite{Vega2005b, *Vega2009, *Vega2011}.
The phase diagrams in Fig.~\ref{fig-phaseDiag-both} are clearly different, but the gradients of the coexistence curves are very similar in both, which may be somewhat surprising given how sensitive phase diagrams can be to small changes in the potential.

\begin{figure}[t]
\centering
\includegraphics{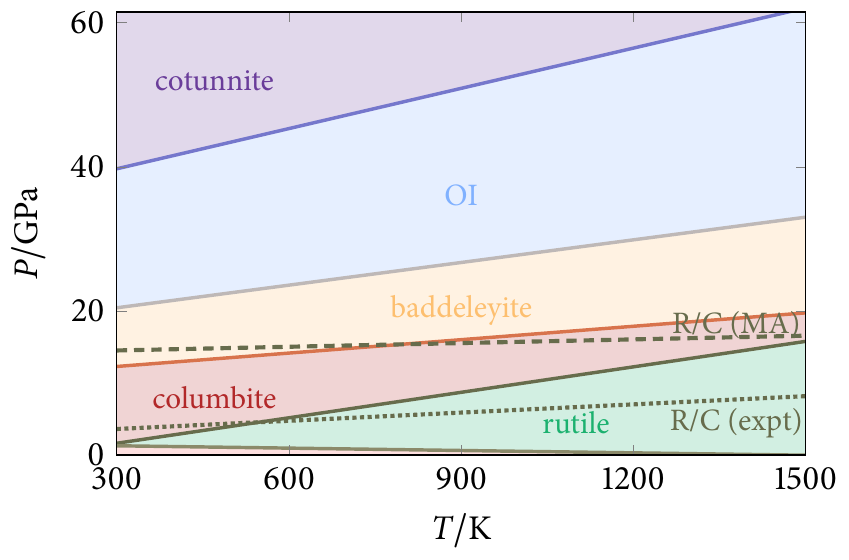}
\caption{Pressure--temperature phase diagram redrawn from the data reported by Mei and co-workers~\cite{Mei2014}. Two further lines are shown corresponding to the rutile/columbite (R/C) transition: the line labelled `expt' corresponds to the experimental data of Ref.~\onlinecite{Akaogi1992}, and the line labelled `MA' corresponds to the metastable transition of the MA potential.}\label{fig-phaseDiag-Mei2014}
\end{figure}

Irrespective of their similarity (or otherwise) to one another, the two phase diagrams shown in Fig.~\ref{fig-phaseDiag-both} look different from the phase diagrams at absolute zero previously computed in DFT-based calculations, such as that determined by Mei and co-workers~\cite{Mei2014}, which is illustrated in Fig.~\ref{fig-phaseDiag-Mei2014}.
Although it is not necessarily immediately obvious which phase diagram is closer to experiment, the large region of stability of the pyrite phase, which has only been considered in theoretical investigations~\cite{Zhu2014} of TiO$_2$, but has not been reported experimentally, suggests that empirical potentials are not particularly good at describing bulk phase behaviour.
In addition, the P\textit{ca}2$_1$ (oxygen-displaced fluorite) phase is missing altogether in the case of the MA potential, even though Fu~\textit{et al.}\ suggest using DFT calculations and the quasi-harmonic approximation that this phase is stable at \SI{35}{\giga\pascal} at temperatures beyond \SI{476}{\kelvin}~\cite{Fu2013}. However, the P\textit{ca}2$_1$ phase was again first proposed by Zhou~\textit{et al.}\ based on a DFT calculation~\cite{Zhou2010}, and it is not yet clear that it is a phase that might be seen in experiment.

The phase boundary between the rutile and the columbite phase has been studied experimentally~\cite{Withers2003, Akaogi1992}, and is also illustrated in Fig.~\ref{fig-phaseDiag-Mei2014}.
This coexistence line was obtained experimentally from mixtures of the two phases to minimise hysteresis effects.
While the DFT-based calculation of Mei and co-workers captures the stability of the columbite phase, the gradient of the coexistence line is perhaps better captured by the MA potential.
In contrast to the experimental findings, the LHZ potential does not exhibit the columbite phase as being thermodynamically stable at all, while for the MA potential, it is enthalpically stable at absolute zero over a small pressure range ($\sim$\SI{0.6}{\giga\pascal}) in between the rutile and the pyrite phases, but remains stable only up to about \SI{20}{\kelvin}.
However, although the columbite phase is not necessarily thermodynamically stable, there is nevertheless a phase transition between the rutile and the columbite phases at approximately \SI{15}{\giga\pascal} for the MA potential (shown in Fig.~\ref{fig-phaseDiag-Mei2014}) and at approximately \SI{6}{\giga\pascal} for the LHZ potential at \SI{700}{\kelvin}.
The fact that a phase transition occurs in experiment does not, of course, mean that either of the phases in question is necessarily thermodynamically stable.
Similar kinetic effects are readily observed in brute-force simulations with empirical potentials.
For example, using both the LHZ and the MA potentials, the baddeleyite phase spontaneously converts to the columbite phase at sufficiently low pressures (below $\sim$\SI{17}{\giga\pascal} for the MA potential and below $\sim$\SI{5}{\giga\pascal} for the LHZ potential).
Columbite's chemical potential is indeed lower than that of baddeleyite at these pressures; however, the pyrite phase is thermodynamically even more stable than columbite, illustrating that kinetics must play an important role.
This is not an atypical example; there are many phase transitions that occur spontaneously to a metastable phase, and in order to be able to say anything about the true thermodynamic stability, free-energy calculations are indispensable.
Determining a `thermodynamic' phase diagram would be a challenging proposition in experiment, but it would be incredibly valuable for the construction of better empirical models.

There is also experimental work on the phase transition between columbite and baddeleyite~\cite{Tang1993, StaunOlsen1999}, predicting a coexistence line of $T/\si{\kelvin}-273=188.7P/\si{\giga\pascal}-2192.5$~\cite{Tang1993}, in reasonable agreement with the quasi-harmonic DFT-based approach of Mei~\textit{et al.}\cite{Mei2014}
At \SI{700}{\kelvin}, the coexistence pressure should be approximately \SI{14}{\giga\pascal}, with baddeleyite more stable at higher pressures.
As can be seen from Fig.~\ref{fig-chemPot-both}, for the MA potential, the baddeleyite phase is never more stable than the columbite phase, while for the LHZ potential, a phase transition occurs at \SI{9.5}{\giga\pascal}.
However, the thermodynamically stable phase for the LHZ potential at this pressure is the pyrite phase, and so neither the LHZ potential nor the MA potential appears to result in a phase diagram that is consistent with the available experimental data.
Finally, although this is not important for a comparison between different simulations, when comparing to experimental data of phases at high temperatures in particular, it is worth noting that the higher the temperature, the higher the intrinsic vacancy concentration is likely to be~\cite{Sarkar2019}, which has significant implications for the physical and chemical properties of TiO$_2$; we have not accounted for this at this stage.

Free-energy calculations and the resulting phase diagrams allow us to account for the role of entropy in controlling phase behaviour.
While enthalpy differences between the phases are sufficient to exclude the majority of high-free-energy phases when computing the phase diagram, they are not sufficient to determine the precise points of thermodynamic stability.
For example, for the MA potential, if only the enthalpy were taken into account, the OI analogue would persist in being thermodynamically stable over the entire temperature range shown in Fig.~\refSub{a}{fig-phaseDiag-both}, with an approximately \SI{6}{\giga\pascal} region of stability even at \SI{1600}{\kelvin}.
Once entropic effects are properly accounted for, the OI analogue is relatively destabilised and ceases to be thermodynamically stable beyond approximately \SI{500}{\kelvin}, illustrating the necessity of accounting for the entropy of the phases in question.

\begin{table}
\caption{Elastic constants $C_{ij}/\si{\giga\pascal}$ of rutile for the MA and LHZ potentials at \SI{1}{\bar} and \SI{0}{\kelvin}. Experimental results correspond to \SI{4}{\kelvin}, except for those given in brackets, which refer to the lowest temperature reported for these quantities, \SI{298}{\kelvin}, and so should be interpreted with caution.}\label{table-elasticConst}
\centering
% \figureversion{lf,tab}
\sisetup{detect-weight=true, detect-family=true, detect-mode=true}%,text-rm={\figureversion{tab,lf}}}
\newcolumntype{Z}{S[table-format=3, table-text-alignment=center]}
\begin{tabular}{l Z Z Z Z Z Z}
\toprule
 & {{$C_{11}$}} & {{$C_{33}$}} & {{$C_{44}$}} & {{$C_{66}$}} & {{$C_{12}$}} & {{$C_{23}$}} \\
Experiment~\cite{Manghnani1972} & 289 & {(484)} & {(124)} &  227  & 197 & {(145)}\\
MA & 322 & 444 & 124 & 226 & 227 & 146 \\
LHZ & 342 & 498 & 122 & 280 & 281 & 177 \\\bottomrule
\end{tabular}
\end{table}

Since the phase diagrams in Fig.~\ref{fig-phaseDiag-both} correspond to pressurisation along the vertical axis, it is reasonable to predict that the densities of the phases should increase, or equivalently that the volume per particle decreases, $\upDelta_\text{trs} v < 0$, for phase transitions that occur as the system is pressurised.
This is indeed the case: typical values of $\upDelta_\text{trs} v$ are between \SI{-0.1}{\angstrom\cubed} and \SI{-0.6}{\angstrom\cubed}, except for the LHZ potential's rutile--pyrite transition, where $\upDelta_\text{trs} v \approx \SI{-1}{\angstrom\cubed}$.
Using the Clapeyron equation [Eq.~\eqref{eqn-clapeyron}] and the relation $\upDelta_\text{trs} H = T_\text{trs}\upDelta_\text{trs} S$, we can determine the entropy change from the gradient of the phase coexistence curve.
We can also compute the entropy difference at the point of coexistence between two phases readily using the free energy and enthalpy calculations that were required to compute the chemical potentials to begin with.
The two approaches to obtaining $\upDelta_\text{trs} S$ agree well.
Although these quantities are quite small in absolute terms (of the order of $\pm0.1k_\text{B}$ per particle), the various transitions have different signs of the entropy, resulting in different signs of the gradient $\der P /\der T$, and this ultimately leads to a significant dependence of the phase behaviour on temperature.
Interestingly, if we compare the empirical potential phase diagrams of Fig.~\ref{fig-phaseDiag-both} to the one obtained from DFT calculations and the quasi-harmonic approximation (Fig.~\ref{fig-phaseDiag-Mei2014}), one striking difference is that the gradient of the OI--cotunnite transition is different, suggesting that despite the decrease in volume per particle, $\upDelta_\text{trs} S > 0$ for empirical potentials, while $\upDelta_\text{trs} S < 0$ in the work of Mei \textit{et al.}
This may be a reflection of the fact that the OI phase as determined in experiment is not stable with either the MA or the LHZ potentials, and spontaneously transforms into a somewhat different polymorph belonging to the same space group (Fig.~\ref{fig-additionalPhases}), illustrating a further possible weakness of the empirical potentials.

Finally, while hysteresis and metastability make true thermodynamic behaviour difficult to determine both in computer simulation and in experiment, there are other properties which are more readily accessible and that can be used to benchmark simulation results to experiment.
In addition to the density, which we considered above, we can determine elastic constants of a crystal~\cite{Matsui1991}.
Elastic properties were used in the original parameterisation of the MA potential, but do not seem to have been considered in later reparameterisations.
At absolute zero, thermal fluctuations vanish and the elastic components can be determined particularly easily simply by shearing the crystal, minimising the energy, and determining the response of the shear tensor to this deformation.\cite{Note4}
For a tetragonal crystal such as rutile, there are six independent non-zero components of the elasticity tensor, and these have been measured experimentally as a function of temperature and pressure~\cite{Manghnani1972,Isaak1998}.
For other crystal phases, such detailed experimental measurements are not yet available~\cite{Swamy2001}.
The elastic constants calculated for rutile using the MA and the LHZ potentials are reported in Table~\ref{table-elasticConst}.
On the whole, the MA potential appears to be slightly more consistent with the experimentally determined elastic constants; however, neither potential performs exceptionally well.
In particular, the Cauchy relation $C_{12}=C_{66}$ for tetragonal crystals is largely satisfied in simulations, but less so in experiment, which suggests that directional (e.g.~covalent) interactions might be significant~\cite{BornHuang1954}, and these are not accounted for by either the MA or the LHZ potential.

\section{Conclusions}

By considering the MA and LHZ potentials and their phase behaviour, we have shown that what may at first glance appear to be trivial changes to interparticle potentials can lead to significant differences in mechanical and thermodynamic behaviour.
The main lesson to learn from this work is that reparameterising a potential is often a more challenging prospect than it might initially seem.
A number of researchers have independently reparameterised the MA potential with a Lennard-Jones form to account for the interparticle repulsion, which on the face of it appears to be a promising approach to take and can be successful around the minimum of the potential energy well.
However, individual parts of a composite potential should not be considered in isolation, and in the case of the MA potential, because of the favourability of the electrostatic interactions between ions of opposite charge, the minimum of the combined potential is far from the minimum of the Buckingham part of the potential.
This results in significantly different densities of phases simulated using the MA potential and the reparameterised potentials.
For any applications where empirical potentials are used to probe only the qualitative behaviour of a substance, this is likely not a significant concern, but if more quantitative insight is needed, particular care must be taken when (re)parameterising potentials.

The main motivation behind a number of reparameterisations of the MA potential has been the fact that the Lennard-Jones potential is more broadly available in ready-made simulation packages, and that many biological molecules have well established interactions with Lennard-Jones-type particles, which are sometimes assumed to be transferable across different systems.
This is a questionable assumption, and of course it is not sufficient merely to reparameterise the MA potential in terms of the LJ potential in order to be able to account for such interactions.
Since TiO$_2$ will interact with biological molecules at interfaces which are likely to be charged in some form, at least at the local level, and which are therefore difficult to model properly, it is not \textit{a priori} obvious that empirical potentials can describe such interactions sufficiently well to capture the behaviours of interest and to be useful in practice.
Nevertheless, for some properties, such as Mie scattering of nanoparticles in aqueous media, it is likely that empirical potentials should be able to capture the fundamentals of the behaviour well enough, as the relevant lengthscale of interest is sufficiently large.
Empirical potentials can also be used to describe bulk behaviour very well if they are appropriately parameterised.

While we have shown that the reparameterised LHZ potential behaves differently from the MA potential itself, it is not the use of the Lennard-Jones potential that is problematic: difficulties arise when we assume that we can replace one repulsive potential with another without considering the overall potential as a whole.
For example, a back-of-the-envelope reparameterisation of the MA potential with a LJ potential such that the minimum of the potential is closer to the minimum of the MA potential for Ti--O interactions exhibits densities of the various polymorphs that are within about \SI{3}{\percent} of the MA densities (as opposed to up to \SI{15}{\percent} for the LHZ potential)~\cite{Note5}.
It should in principle be possible to design a simple empirical potential for titanium dioxide using either a Lennard-Jones potential or another type of repulsive potential altogether, but the choice of all the parameters of the overall potential should ideally be based on experimental data.
As has been demonstrated with empirical water models~\cite{Vega2005b, *Vega2009, *Vega2011}, computing phase diagrams and comparing coexistence lines with those that can be obtained experimentally can be a very helpful way of parameterising good empirical potentials.

In this work, we have used free-energy calculations and classical statistical mechanics without any (nuclear) quantum corrections to compute phase diagrams~\cite{Note6}. This almost certainly means that the low-temperature behaviour that we see is not correct.
By contrast, several DFT-based approaches have been combined with the (quasi-)harmonic approximation of the phonon modes to estimate phase diagrams for titanium dioxide.
This approach can only capture the entropy around the potential energy minimum, and so the expectation is that such phase diagrams might not be reliable at high temperatures.
While density-functional theory predictions depend significantly on the functional approximation employed, it does seem that at least some previous DFT-based work results in phase behaviour that is closer to experiment~\cite{Mei2014}, but various DFT-based approaches can give vastly different predictions~\cite{Luo2016}.
Moreover, since experimental data are relatively patchy, and it is difficult to know which phase transitions are between stable and which are between metastable phases in experiment, it would be particularly exciting to see further experimental work on determining the full thermodynamic phase diagram of TiO$_2$, which would enable us to benchmark theoretical and computational predictions against experimental data with more confidence.
However, the fact that certain phases are not as stable with empirical potentials as they appear in experiment indicates that there is scope for improving empirical pair potentials.
Encouragingly, since the MA and LHZ potentials are capable of simulating nearly every polymorph of TiO$_2$ suggested in experiment or by DFT calculations thus far, it is certainly not unreasonable to venture that simple empirical pair potentials can be devised that will be able to capture the whole range of crystalline polymorphs of the material \textit{and} be consistent with experimental data available.

Empirical potentials such as the ones considered here can be very useful in practice when investigating phenomena that do not depend too strongly on surface properties, and using simple models allows us to learn about the fundamentals of the physics of a given system.
However, more complex models will be necessary to describe many of the interesting features of titanium dioxide.
It has recently been shown that machine-learned potentials based on DFT calculations can be used to predict with surprising ease the thermodynamic behaviour of water~\cite{Cheng2019}; given the interest in titanium dioxide, it may be particularly helpful to investigate whether such an approach may also be fruitful.
\textit{Ab initio} molecular dynamics simulations to investigate the structure of amorphous and crystalline TiO$_2$ have already been undertaken~\cite{Mavracic2018}, and despite the intrinsic difficulty of treating long-ranged electrostatics with machine-learned potentials~\cite{Bartok2015}, an accurate potential for TiO$_2$ that goes beyond empirical pair potentials would be particularly useful to develop~\cite{Note7}.
For both empirical and machine-learned potentials, however, free-energy calculations to compute the phase diagram can help to give us confidence in the potential's performance and predictive power.

\begin{acknowledgments}
I would like to thank Daan Frenkel for many very insightful discussions.

Supporting data are available at the University of Cambridge Data Repository, \href{https://doi.org/10.17863/cam.41537}{doi:10.17863/cam.41537}.
\end{acknowledgments}

% \bibliography{bibliography}
% \bibliographystyle{aipnum4-1}


\end{document}